\documentstyle[rotate,epsf]{mn}

\title[Soft X-ray transient light curves as standard candles: exponential 
versus linear decays]
{Soft X-ray transient light curves as standard candles: exponential 
versus linear decays}
\author[T.~Shahbaz, P.A.~Charles and A.R.~King]{
T.~Shahbaz$^{1}$, P.A. Charles$^{1}$ and A.R.~King$^{2}$  \\
$^{1}$University of Oxford, Department of Astrophysics, Nuclear Physics
Building, Keble Road, Oxford, OX1~3RH, UK \\
$^{2}$ Astronomy Group, University of Leicester,
Leicester, LE1~7RH, UK}

\begin{document}

\maketitle

\begin{abstract}

\noindent
A recent paper by King \& Ritter (1998: hereafter KR) proposed that the light
curves of soft X--ray transients (SXTs) are dominated by the effect
of irradiation of the accretion disc by the central X--rays. This
prevents the onset of the cooling wave which would otherwise return
the disc to the quiescent state, and so prolongs the outbursts beyond
those in dwarf nova discs. KR show
that the decay of the resulting X--ray light curve should be
exponential or linear depending
on whether or not the observed peak X--ray luminosity is sufficient to
ionize the outer edge of the accretion disc. 
Here we examine the observed X--ray decays, and show that they
are exponential or linear according as the peak luminosity is greater or
smaller than the critical value defined by KR, strongly suggesting
that the light curves are indeed irradiation--dominated. We show
further that the occurrence of an exponential or linear decay tends to
favour the same type of decay in subsequent outbursts, so that systems
usually show only one or the other type.
We use the equations of KR
and the observed X--ray light curve to determine the size $R_{h}$ of
the hot disc at the peak of the outburst. For exponential
decays, $R_{h}$ is found to be comparable to the circularization
radius, as expected since the disc consists entirely of material
transferred from the secondary since
the previous outburst. Further, $R_h$ is
directly proportional to the time at which one sees the
secondary maximum ($t_{s}$),
as expected if $t_{s}$ is the viscous timescale of the irradiated disc.
This implies that the orders of magnitude of the viscosity parameter
$\alpha$ and disc aspect ratio $H/R$ are such that $\alpha (H/R) \sim
0.01$. Observation of a secondary maximum calibrates the peak luminosity
and gives the distance ($D_{\rm kpc}$) to the source as

\begin{displaymath}
D_{\rm kpc} = 4.3 \times 10^{-5} t_{s}^{3/2} \eta^{1/2} f^{1/2} F_{p}^{-1/2}
\tau_{d}^{-1/2}
\end{displaymath}

\noindent
where $F_{p}$ is the peak flux, $\tau_{d}$ is the $e$--folding time of
the decay in days, $\eta$ is the radiation efficiency parameter and $f$
is the ratio of the disc mass at the start of the outburst to the maximum
possible.

\end{abstract}

\section{Introduction}

The soft X-ray transients (SXTs) are a subclass of the low--mass X--ray
binaries in which a low--mass star transfers material to a neutron star or
black hole. They undergo large outbursts, reaching X--ray luminosities of
order the Eddington limit (see Tanaka \& Lewin 1995 and Tanaka \& Shibazaki
1996 for reviews). The outbursts have similarities to those of dwarf novae,
but with important differences, one being that the timescales are much
longer; a dwarf nova outburst lasts a few days and typically recurs every few
weeks, whereas in SXTs the corresponding timescales are months and years. It
is generally accepted that the dwarf nova outburst results from a
thermal--viscous instability in the accretion disc (see Cannizzo 1993 for a
review). To some extent the models used for dwarf novae can reproduce the
correct timescales for the SXT outburst, but only by reducing the viscosity
parameter by a factor of 10--100 and choosing a particular functional form
for it (e.g. Cannizzo, Chen \& Livio, 1995).

Although there are similarities between the discs during outburst in dwarf
novae and SXTs, an important observed difference is that the discs in the
SXTs are heavily irradiated by the central X--ray source (van Paradijs
\& McClintock 1994; Shahbaz \& Kuulkers 1998). Recently
King \& Ritter (1998) have pointed out that irradiation prolongs the
entire SXT outburst cycle, forcing SXT light curves to have long
exponential or linear tails, because the outburst can only shut off
through the viscous decay of the central accretion, rather than via a
cooling front as is the case in dwarf novae. (For counterarguments see
Cannizzo 1998.) In this paper we apply some of the key features of the KR
irradiation model to the X--ray light curves of SXTs.

\section{Outbursts of irradiated discs}

KR show that the light curves of SXTs can be explained by a disc
instability model, modified to take account of irradiation by the central
X--ray source during the outburst. Irradiation prevents the disc
returning to the cool state until the central accretion rate is sharply
reduced. If irradiation is strong enough to ionize all of the disc out to
its edge, KR show that the X--ray light curve will be a roughly
exponential decay, with most of the disc mass being accreted on a viscous
timescale. The recurrence time to the next outburst will be long, as the
disc has to be rebuilt by mass transfer from the companion star. If
instead the X-rays are too weak to ionize the whole disc, KR show that
the X--ray light curve should be a roughly linear decline, returning to
quiescence after a viscous timescale. In this case much of the disc mass
is not accreted, allowing the next outburst to begin after a shorter
recurrence time than if there had been an exponential X--ray decay.

\subsection{The critical luminosity}

KR give expressions for the critical central accretion rates required
to ionize the disc out to a given radius $R_h$. 
There are two rates, depending on
whether the central source is point--like or disc--like. In the second
case irradiation is weaker by a typical factor $H/R$, 
since a disc--like source is foreshortened by this factor as seen by matter at
the disc surface. As a result the critical accretion rate (and thus
luminosity) is higher by a factor $\sim R/H$ than in the point--like case.
The critical rates (14) and (15) of KR are equivalent to critical luminosities

\begin{equation}
L_{\rm crit}(\rm point$--$\rm like) =  
3.7\times 10^{36} R_{11}^{2}~~~ {\rm ergs~s}^{-1} \label{pt}
\end{equation}

\begin{equation}
L_{\rm crit}(\rm disc$--$\rm like) =  
1.7\times 10^{37} R_{11}^{2}~~~{\rm ergs~s}^{-1} \label{disc}
\end{equation}

\noindent
Here $R_{11}$ is the ionized disc radius $R_h$ in units of $10^{11}$~cm.
The maximum possible value of the outer disc radius, and hence the radius
$R_h$ of the ionized region, is given by the tidal radius $R_{\rm T}$ of
the disc (we use 80 per cent of the Roche lobe of the primary in the
applications in this paper, but it should be noted that the fraction can
vary between 70 and 90 per cent). If the accretor is a neutron star it
must behave as a point--like source, as its surface will be heated by
accretion. If the accretor is a black hole it is possible for it to
behave as a disc--like accretor (the relevant formulae were labelled `NS',
`BH' in KR). Indeed this effect probably accounts for the very strong
tendency of black--hole systems with low--mass companion stars to be
transient (King, Kolb \& Szuszkiewicz, 1997): the {\it mean} mass
transfer rate $-\dot M_2$ lies below $L_{\rm crit}$(disc--like) for
all black--hole systems, almost irrespective of the nature of the
companion star, so that a persistent state with a stably irradiated disc
is impossible for them. However, during outbursts, all transients develop
a hard power--law X--ray component which is very probably a central
X-ray corona, in addition to the softer blackbody--like component which
probably comes from the disc. Thus it is likely that even black--hole
accretors become point--like X--ray sources during SXT outbursts, making
$L_{\rm crit}$(point--like) given by (\ref{pt}) the correct formula to
use during the outburst. This is the procedure adopted in this paper
(see Fig. 1). Note that the use of $L_{\rm crit}$(point--like) during
outbursts and $L_{\rm crit}$(disc--like) for the mean behaviour is
consistent, provided that the power--law coronal component becomes weaker
than the blackbody component once the central accretion rate $\dot M_c$
declines below the mean mass transfer rate $-\dot M_2$ near the end of an
outburst.

Equipped with the appropriate value of $L_{\rm crit}$, we can now
predict the form of the light curve decay.
If the observed peak X--ray luminosity is less than the value of $L_{\rm
crit}$ corresponding to its outer radius, the outer edge of the disc is
not strongly irradiated and the decay will be linear. If however the
observed luminosity is greater than this critical luminosity, it is
strong enough to ionize the outer parts of the disc and the decay will
initially be exponential. Both kinds of decay are observed.

\subsection{Linear decays}

In a linear decay the X--ray luminosity obeys

\begin{equation}
L_X = \eta c^2\biggl({3\nu\over B_1}\biggr)^{1/2}
\biggl[M_{\rm h}^{1/2}(0) - 
\biggl({3\nu\over B_1}\biggr)^{1/2}t\biggr]. \label{lin}
\end{equation}

\noindent
(using eqn 23 of KR), where the decay started at time $t=0$. Here $\nu$
is the kinematic viscosity near the edge of the heating front, $M_h(0)$
is the heated mass at the start of the outburst (peak luminosity), and

\begin{equation}
B_1  \simeq 4\times 10^5\ {\rm (cgs)}.
\label{eq:B}
\end{equation}

\noindent
Thus the peak value $L_p$ of $L_X$ is given by setting $t=0$, and
gives

\begin{equation}
M_h(0) = {B_1L_p^2\over 3\nu\eta^2 c^4}
\end{equation}

\noindent
Defining $t_{1/2}$ as the time for $L_X$ to fall by a factor 2 we get

\begin{equation}
\nu = {B_1L_p\over 6\eta t_{1/2}}
\end{equation}

\noindent
Combining these two relations and measuring $t_{1/2}$ in days we find

\begin{equation}
M_h(0) = {2L_pt_{1/2}\over \eta c^2} = 
1.9 \times 10^{-10}L_p\eta^{-1}t_{1/2}~~~{\rm g}
\label{ml}
\end{equation}

\noindent
and

\begin{equation}
\nu = 2.14\times 10^{-27} B_1 L_{p} \eta^{-1} t_{1/2}^{-1} 
~~~ {\rm cm^{2}~s}^{-1}.
\end{equation}

\noindent
Thus given the peak X--ray luminosity and the time taken for the flux to
drop by a factor 2 we can determine the ionized disc mass at the start of
the outburst and the disc viscosity. Since for a linear decay $L_X$ lies
below the critical value which would ionize the whole disc (by
definition), the radius of the ionized disc at the start of the outburst
is given by

\begin{equation}
R_h(0)^2 = {B_1L_p\over \eta c^2}
\end{equation}

\indent
A linear decay in an SXT implies that the ratio $L_X/$(outer disc
radius)$^2$ is smaller than a critical value. This can happen either
because (a) the central X--rays are weaker than usual, or (b) the disc is
very extended, which in turn requires the tidal radius $R_{\rm T}$ to
allow a large disc radius. Case (a) is seen in Aql X--1 (see section
4.1); the 19--hr period means a relatively small disc, but the X--ray
luminosity only reaches peak values $L_p < 10^{36}$~erg~s$^{-1}$. Case
(b) must apply in sufficiently long--period binaries: the disc cannot be
smaller than the circularization radius, which increases with orbital
period, while the X--ray luminosity cannot greatly exceed the Eddington
limit. An example is GRS 1744--63, (P = 11.8 days; the longest period of
any SXT) which is linear right from the start of the observed outburst
decline ($\dot{M_{c}}=1.5\times 10^{-8}~M_{\odot}~{\rm yr}^{-1}$ cf
$\dot{M_{\rm crit}}({\rm point-like})
=2.2\times 10^{-8}~M_{\odot}~{\rm yr}^{-1}$).

\subsection{Exponential decays}

For systems in which the peak X--ray luminosity is strong enough to keep the
whole disc in a hot state, KR show that the X--rays decrease
exponentially, i.e.

\begin{equation}
L_X = \eta c^2 R_h\nu\rho f\exp(-t/\tau),
\label{exp}
\end{equation}

\noindent
where $\rho \simeq 3\times 10^{-8}$~g~cm$^{-3}$, $f$ is the ratio of
the disc mass at the start of the outburst to the maximum possible, and

\begin{equation}
\tau = {R_h^2\over 3\nu}.
\label{tau}
\end{equation}

\noindent
We can fit the form of (\ref{exp}) at $t$=0 to the observed X--ray light
curve, to give the normalisation of the light curve and the decay time
allows one to determine $\nu$ and $R_h$. From (\ref{exp}) at $t$=0 we get

\begin{equation}
L_p = \eta c^2 R_h\nu\rho f,
\label{p}
\end{equation}

\noindent
and combining this equation with (\ref{tau}) gives 

\begin{equation}
\nu = 1.68\times10^{-11} L_{p}^{2/3} \eta^{-2/3}f^{-2/3} \tau_{d}^{-1/3}
~~~ {\rm cm^{2}~s}^{-1}
\end{equation}

\noindent
and

\begin{equation}
R_h = 2.09 \times 10^{-3} L_{p}^{1/3} \eta^{-1/3}f^{-1/3} \tau_{d}^{1/3}
~~~ {\rm cm}
\label{rad} 
\end{equation}

\noindent
where $\tau_d$ is $\tau$ measured in days.

\indent
Equation (\ref{exp}) is equivalent to the statement that the hot disc
mass $M_h$ decays exponentially on the viscous timescale $\tau$, i.e.

\begin{equation}
L_X = \eta c^2 {M_h\over \tau},
\end{equation}

\noindent
(cf eqn 4 of KR), so the total heated mass at the start of the
outburst is

\begin{equation}
M_{h}(0) = {L_p\tau\over \eta c^2} 
= 1.62 \times 10^{-16} L_{p} \eta^{-1} \tau_{d} ~~~{\rm g}
\label{mh}
\end{equation}

\noindent
Note that this equation is very similar to equation (\ref{ml}), i.e. the
mass of the hot disc in the linear case. Combining the first form of this
equation with equations (\ref{p}) and (\ref{tau}) shows that the heated
mass at the start of the outburst is

\begin{equation}
M_h(0) = {\rho f R_h^3\over 3}
\end{equation}

\noindent
The maximum possible value of this mass is given by assuming that the
disc has the critical density everywhere ($f=1$), and that its outer
radius has the largest possible size, i.e. $R_h = R_{\rm T}$. This gives 

\begin{equation}
M_{\rm max} = {\rho R_{\rm T}^{3} \over 3},
\label{mmax}
\end{equation}

\noindent
where $\rho\simeq 
3.0\times 10^{-8}$~g~cm$^{-3}$ and $R_{\rm T}$ the tidal radius.
Since the latter quantity is a function of binary parameters, this
value can be compared with the observational estimate (\ref{mh}).

\begin{table*}
\caption{Parameters for the SXTs}
\begin{center}
\begin{tabular}{lcccccccc}
Source &  Type & Year & log $F_{p}$ & D & log $L_{p}$ &
       $\tau_{d}$ & $P_{orb}$ & $R_{\rm T}$ \\
&  &  &  (ergs~s$^{-1}$~cm$^{-2}$)  & (kpc) &
   (ergs~s$^{-1}$) & (days) & (hrs) & (10$^{11}$~cm) \\
& & & & & & & &  \\
SAX~J1808.4--3658 & E & 4/1998 & -8.62 & 4.0    & 36.66 & 8.4  & 2.0  & 0.2 \\
GRO~J0422+32 & E & 7/1992 & -7.46 & 2.0$\pm$0.5 & 37.22 & 40.1 & 5.1  & 0.9\\
A0620--00    & E & 8/1975 & -6.00 & 0.9$\pm$0.4 & 37.99 & 26.3 & 7.8  & 1.2\\
GS~2000+25   & E & 5/1988 & -6.71 & 2.0$\pm$1.0 & 37.97 & 30.1 & 8.3  & 1.3\\
GS~1124--683 & E & 1/1987 & -6.77 & 4.0$\pm$1.0 & 38.51 & 28.3 & 10.4 & 1.3\\
Cen X--4     & E & 5/1979 & -7.11 & 1.2$\pm$0.3 & 37.13 & 4.8  & 15.1 & 1.1\\
Aql X--1     & L/E & 5/1978 & -7.50 & 2.3$\pm$0.1 & 37.30 & 32.6/35.7 
& 19.0 & 1.3 \\
Aql X--1     & L & 2/1997 & -8.08 & 2.3$\pm$0.1 & 35.72 & 15.0 & 19.0 & 1.3\\
Aql X--1     & L & 8/1997 & -8.28 & 2.3$\pm$0.1 & 35.52 & 15.0 & 19.0 & 1.3\\
4U~1543--47  & E & 3/1971 & -7.24 & 8.0$\pm$1.0 & 38.65 & 42.7 & 29.6 & 2.3\\
GRO~J1655--40 & L & 6/1997 & -7.40 & 3.2$\pm$0.2 & 37.69 & 30.0 & 62.9 & 4.4  \\
GRO~J1744--28 & L & 1/1997 & -7.72 & 6.5$\pm$1.5 & 37.98 & 30.0 & 283.2 & 7.6 \\
\end{tabular}
\end{center}
     
\begin{tabular}{l}
E and L refer to exponential and linear decays respectively \\
$F_{p}$, the peak X-ray flux and $\tau_{d}$ are taken 
from Chen, Shrader \& Livio (1997) or the $RXTE$ light curves. \\
We have assumed total binary masses 2$M_{\odot}$ and 
10$M_{\odot}$ for the neutron--star and black--hole systems respectively. \\
For the linear decays $\tau_{d}$ is $t_{1/2}$. \\
The distance to Aql X--1 is taken from Shahbaz et al. (1998).
\end{tabular}
\end{table*}

\section{Conditions in the quiescent disc}

We have seen that exponential decays are expected when the whole of
the accretion disc can be kept in the hot state by the central
X--rays. In this case a substantial fraction of the disc
mass is likely to to be accreted in the course of the outburst. This
in turn suggests a substantial recurrence time before the disc can be
rebuilt and produce the next outburst. By contrast, in a linear decay,
the outer disc remains cool and is essentially unaffected by the
outburst. Since it probably contains a large fraction of the disc mass,
viscous evolution can produce the critical density again somewhere within the
disc fairly quickly, allowing another outburst. Usually, such repeated
linear outbursts can be expected to deplete the disc mass faster than
it is refilled by mass transfer. Thus after a series of linear
outbursts, the disc may well enter a much longer quiescent state in
which the disc is rebuilt. This kind of `supercycle', with a long
quiescent state alternating with a rapid series of linear outbursts,
is reminiscent
of the outburst behaviour actually observed in linear--decline systems
such as GRO~J1744--28: no outbursts at all were seen by any instrument
until the first observed one in 1996, yet this was followed by a
further outburst only $\sim 1$~yr later (Kouveliotou \& van Paradijs 1998).

A more subtle point concerns the size of the rebuilt disc after an exponential
decay. Since this disc consists almost entirely of material
transferred from the secondary star since the last outburst, and there
has been negligible accretion on to the central star in this time, its
mean specific angular momentum $j$ must be precisely the same as that
transferred through the inner Lagrange point $L_1$, i.e.
\begin{equation}
j = (GM_1R_{\rm circ})^{1/2},
\label{j}
\end{equation}
where $M_1$ is the primary mass and
$R_{\rm circ}$ is by definition the circularization radius,
i.e. the radius of the Keplerian circle having the same specific angular
momentum as $L_1$: typically $R_{\rm circ} \sim 0.5R_{\rm T}$.
[Eqn \ref{j} does {\it not} hold for a disc which has
undergone accretion on to the central star, as it increases $j$
above $(GM_1R_{\rm circ})^{1/2}$ by selectively ridding itself of
low--angular--momentum material.] Thus given the surface density 
$\Sigma (R)$ of a post--exponential disc, the condition
(\ref{j}) fixes its outer edge $R_{\rm out} [=R_h(0)$ when the
outburst starts] through the requirement
\begin{equation}
{2\pi\int_0^{R_{\rm out}}(GM_1R)^{1/2}\Sigma R{\rm d}R\over 
2\pi\int_0^{R_{\rm out}}\Sigma R{\rm d}R} = (GM_1R_{\rm circ})^{1/2}.
\end{equation}
With $\Sigma \propto R^n$ we find
\begin{equation}
R_{\rm out} = \biggl({2n + 5\over 2n+4}\biggr)^2R_{\rm circ}.
\label{rout}
\end{equation}
A quiescent disc generally has $n \sim 1$, giving 
$R_{\rm out} \simeq 1.36R_{\rm circ} \sim 0.7R_{\rm T}$. 
Thus we expect a rather small disc at
the beginning of the next outburst after an exponential decay. Once
the outburst begins the disc will try to evolve viscously to a more
centrally--condensed state ($n <0$), and so will tend to spread
towards $R_{\rm T}$ in order to conserve angular momentum.
However this only occurs when the outburst is well
advanced, and the values of $R_h(0)$ we find for exponential decays
should be significantly smaller than $R_{\rm T}$.
Of course for a disc left over after a linear decay we would
expect a radius comparable with $R_{\rm T}$ in any case. This bimodal
behaviour means that each type of outburst tends to produce
conditions favourable for the same type of decay in the following
outburst. Thus an exponential decay will mean a small disc, which is
more easily kept fully irradiated, making an exponential decay likely
in the next outburst. Similarly, a linear decay leaves behind a large
disc, tending to keep subsequent decays linear also.

\section{Application to the SXTs}

In this section we consider the decline from outburst of a sample of
SXTs. The X--ray light curves which most reflect the bolometric luminosity
lie in the 0.4--10 keV range (see Chen, Shrader \& Livio 1997,
hereafter CSL).  Therefore we only consider systems with X--ray light
curves in this energy range. Systems where only very hard X--ray light
curves are available are more difficult to interpret e.g. GRO~J0422+32
and GRS~1009--45. Also it should be noted that the value for the
observed luminosity depends on the distance to the source. This will
introduce some uncertainty, in particular in the poorly studied or
faint systems where reliable distances are not available. We further
restrict our sample to SXTs with known orbital periods, as this
information is required in order to calculate the tidal and
circularization radii and thus predict the critical
X--ray luminosity needed to keep the outer disc edge ionized. In Table 1
we give the peak X-ray luminosities and decay timescales for the SXTS which
have well sampled X-ray light curves and for which the
distance and orbital period is known.

Fig. 1 shows the critical luminosity ($L_{\rm crit}$) for point--like
central sources (eqn \ref{pt}) plotted against orbital period for
total binary masses $M = 2M_{\odot}, M = 10M_{\odot}$, typical for
neutron--star and black--hole systems respectively.
In each case there are two lines,
corresponding to the fact that the disc radii are typically $R_{\rm
T}$ and $0.7R_{\rm circ}$ for exponential and linear
decays respectively. Accordingly there is a luminosity band in which
exponential and linear decays can coexist, even for the same assumed
type of central X--ray source. In Fig. 1 we assume that the disc
radius is a factor 0.7 smaller in the exponential case compared with
the linear one, so that this luminosity band is just a factor $1/0.7^2
\sim 2$.
If the peak observed luminosity (in the 0.4--10 keV range) is larger than
the relevant value of
$L_{\rm crit}$, the initial decline of the outburst is expected to be
exponential, becoming linear as the luminosity declines through $L_{\rm crit}$.
Outbursts in which $L_{p}$ is less than the relevant 
$L_{\rm crit}$ should be entirely
linear (see KR). We also show in Fig. 1 the observed peak X--ray luminosity
for those SXT outbursts where we can determine the shape of the decline. 
The data are taken from CSL and the $RXTE$ database.
As expected, the outbursts divide into exponential and linear for peak
luminosities above and below the relevant $L_{\rm crit}$.

In Table 2 we determine the physical parameters for accretion discs
in SXTs with known orbital periods and where the shape of the
outburst decline is known.

\subsection{Aql X--1 (=4U1908+005) }

The (February 1997 and August 1997) outbursts of Aql X--1 both had peak
luminosities much less than $L_{\rm crit}$ for a neutron star. Thus we
expect the decline of these outbursts to be entirely linear; as
demonstrated by Fig. 2. However, the 1978 outburst of Aql X--1 (Charles et
al. 1980) is expected to be exponential, since it lies above $L_{\rm
crit}$ for a neutron star. CSL interpret this outburst as having an
initial plateau i.e. its initial decay is exponential. Note that this
would be consistent with the model of KR. It should also be noted that
the X--ray light curve of the 1978 outburst is best fitted with a linear
decline (see Fig. 2), rather than the commonly assumed exponential decay.
However, the uncertainties in the data are large and so the true form of
the decay may be masked.

For the 1997 outbursts of Aql X--1 the values of $M_h(0)$ (2.3 $\times
10^{23}$~g and 1.5 $\times 10^{23}$~g for the February and August
outbursts respectively) are much smaller than one would infer from
assuming that the disc had almost the critical surface density all the
way out to $R_{\rm T}$ [5.5 $\times 10^{24}$~g and 2.9 $\times 10^{24}$~g for
the February and August outbursts respectively; the maximum possible hot
disc mass is given by equation \ref{mmax}]. This may imply that not much of the
mass in the disc is accreted, and so there is sufficient mass left in the
disc after the outburst, such that it may be triggered again much sooner
than otherwise expected. Alternatively, it may be that the disc simply
never reached its maximum mass before the surface density reached its
critical value and triggered an outburst. However, the second possibility
would not result in an ``outside--in'' outburst (which requires a high
disc mass), as is observed in the 1997 August outburst (Shahbaz et al.
1998). The first possibility seems to be consistent with the short
observed recurrence time; Aql X--1 shows quasi--periodic outbursts every
309 and 125 days (Kitamoto, Tsunemi \& Miyamoto 1993). If we take the
recurrence time to be 0.6 yr and use equation (35) of KR for linear
decays, we obtain a mass transfer rate $\sim 2\times 10^{-10}$
$M_{\odot}~{\rm yr}^{-1}$. Note that the estimate of the mass transfer
rate is roughly what is expected for a subgiant filling its Roche lobe in
a binary of this period (see King, Kolb \& Burderi 1996 and CSL).

\subsection{Cen X--4 (=4U1456--32) }

For Cen X--4 with an orbital period of 15 hrs we find $L_{\rm crit}=1.6
\times 10^{37}$~erg~s$^{-1}$ (using eqn. \ref{pt}). Both the 1969 and 1979
outbursts (Evans et al. 1970; Kaluzienski et al. 1980) had peak luminosities
higher than $L_{\rm crit}$ for a neutron star, so one expects exponential
decays. In Fig. 3 we show exponential fits to the decline of the 1969 and
1979 outbursts of Cen X--4, exactly as expected (see Fig. 1).

\subsection{4U1543--47}

The 1971 March X-ray outburst light curve of XN 1543--47 is best fitted
with an initial exponential decay (see Fig. 3; the final part of the decay
may be linear but it is much harder to determine because of the presence
of the secondary maxima. $L_{p}(=4.5\times 10^{38}$~erg~s$^{-1}$)
is higher than $L_{\rm crit}$(point-like) for a 8M$_{\odot}$ binary (we
have used a distance of 8 kpc; see Orosz et al. 1997) and so the compact
object is most probably a black hole.

\begin{table*}
\caption{Derived physical parameters for the SXT accretion discs}
\begin{center}
\begin{tabular}{lcccccccc}
Source & & $t_s$ & $\nu$ & $M_{h}$(0) & $R_{h}$(0) & $c_{S}$ &
     $f^{1/3}R_{h}(0)/R_{\rm T}$ & $M_{h}(0)/M_{max}$ \\
 & & (days)    & (10$^{15}$~cm$^{2}$~s$^{-1}$) & (10$^{23}$~g)  & (10$^{10}$~cm)
&  (10$^{6}$~cm~s$^{-1}$)  &  (\%) & (\%) \\
& & & & & & \\
SAX~J1808.4--3658 & E & 13 & 0.1 & 0.24  & 1.3         & 1.2 & 57  & 18.4 \\
GRO~J0422+32 & E & 38 & 0.2  &   8.3   &  4.4$\pm$0.7  & 1.1 & 50  & 13  \\
A0620--00    & E & 54 & 0.6  &  24.7   &  6.3$\pm$1.1  & 1.2 & 53  & 15  \\
GS~2000+25   & E & 75 & 0.5  &  27.0   &  6.5$\pm$2.1  & 1.2 & 51  & 13  \\
GS~1124--683 & E & 74 & 1.3  &  87.9   &  9.6$\pm$1.6  & 1.4 & 73  & 40  \\
Cen X--4     & E & 12 & 0.2  &   0.4   &  1.6$\pm$0.3  & 1.0 & 15  & 0.3 \\
Aql X--1 5/1978 & L & -  & 3.5  &   8.3   & 24.3$\pm$2.4 & 1.0 & 193  & 4.2 \\
Aql X--1 5/1978 & E & -  & 0.1  &   4.6   & 3.6$\pm$0.4  & 1.0 &  28  & 2.3 \\
Aql X--1 2/1997 & L & -  & 2.0  &   1.0   & 12.5$\pm$1.2  & 0.8 & 99  & 0.5 \\
Aql X--1 8/1997 & L & 18 & 1.3  &   0.6   &  9.9$\pm$1.0  & 0.8 & 79  & 0.3 \\
4U~1543--47  & E & 15 & 1.4  & 183.1   & 12.2$\pm$1.0  & 1.3 & 53  & 15  \\
GRO~J1655--40 & L & 40 & 9.3  &  42.3   & 46.7$\pm$1.9 & 0.8 & 105 & 0.5 \\
GRO~J1744--28 & L & 52 & 17.1 &  39.1   & 53.2$\pm$10.6 & 0.8 & 70  & 0.1 \\
\end{tabular}
\end{center}

\begin{tabular}{l}
For GRO~J0422+32 we use the UV secondary maxima seen after the
BATSE peak (Shrader et al. 1994). \\ 
For GRO~J1744-28 a secondary maximum is seen
in the BATSE data (Kouveliotou \& van Paradijs 1998).
\end{tabular}

\end{table*}

\section{The secondary maximum; a standard candle}

The monotonic decay of X--ray/optical light curves of many of the SXTs
are often interrupted by a flux increase of a factor of 2 or more over a
timescale $\la 10$~d, followed by the resumption of the normal decay. The
secondary maximum has predominantly been seen in the BH SXTs. However, it
should be noted that a closer inspection of the X--ray light curves of Cen
X--4 (Kaluzienski et al. 1980), Aql X--1 (see Fig. 3) and GRO~J1744-28
(Kouveliotou \& van Paradijs 1998) also show similar secondary maxima.
Note that GRO~J1655-40 is the only BH SXT that shows a linear decay (see
Fig. 4).

We define the time in days after the outburst peak that the secondary
maxima occurs as $t_{s}$.  As KR point out, these secondary maxima
features may be caused by material in the outer disc triggering small
outbursts during the linear decay phase.  A related idea is simply that
the increase in viscosity when irradiation begins is more marked in the
outer regions than further in, causing a `pulse' of extra mass to move
inwards. In both cases the secondary maximum should appear one
irradiated--state viscous time after the initial outburst.  In Table 2 we
determine $t_{s}$ and $R_{h}(0)$ for those SXTs in which the secondary
maximum has been observed, which we plot in Fig. 5. Note that all data
points from exponential decays lie in a straight line, whereas the data
from linear decays are systematically higher. This may suggest that there
is a different relation for linear decays, which may not be so surprising
as one would expect the density wave to behave differently depending on
whether or not it started in the outermost part of the original disc.
Only more observations of linear decays with secondary maxima will
resolve this.

A least squares linear fit to the data points from exponential decays gives 

\begin{equation}
R_{h} = 0.128(\pm0.007)10^{10} t_{s} ~~~ {\rm cm};
\label{fit}
\end{equation}

\noindent
where the fit has a correlation coefficient of 0.92 and 1--$\sigma$
errors are given (a constant term in the fit is only significant at the
34 percent confidence level). i.e. this fit is consistent with the
physical requirement that it pass through the origin. Thus given a
complete X--ray outburst of a SXT then $t_{s}$ is determined by the
observation of a secondary maximum. This can then be used in eqn. (\ref{rad})
to predict the peak luminosity of the outburst. The distance to the
source ($D_{kpc}$) can then be determined given the observed peak X--ray
flux $F_{p}$. Thus for exponential decays we obtain

\begin{displaymath}
D_{\rm kpc} = 4.3 \times 10^{-5} t_{s}^{3/2} \eta^{1/2} f^{1/2} F_{p}^{-1/2}
\tau_{d}^{-1/2}
\label{d}
\end{displaymath}

\subsection{The viscosity}

The gradient of the $R_{h}$ versus $t_{s}$ fit is related to the rate
with which $R_{h}$ decreases, i.e. the speed of the density wave
producing the secondary maxima. The speed of this wave is given by the
radial drift velocity

\begin{equation}
V_{r}=\frac{ 3\nu }{2R}=   \frac{3}{2} \alpha c_{S} \frac{H}{R} 
\end{equation}

\noindent
where $\alpha$ is the accretion disc viscosity parameter, $c_{S}$ is the
equatorial--plane sound speed in the hot gas and $H$ is the disc scale
height at disc radius $R$. For an irradiated disc the 
temperature at the edge of the density wave is given by

\begin{equation}
\rm \log T = -4.79 + 0.25\log L_{p} -0.5\log R_{h} ~~~ {\rm K}
\end{equation}

\noindent
(van Paradijs 1996) which then gives $c_{S}$.  We find a 
mean $c_{S}$ of 1.0$\times10^{6}$~cm~s$^{-1}$ (see Table 2), in line
with our expectation that the outer region of the disc must be close
to but just above the ionization temperature $\sim 6500 - 10,000$~K.
>From the gradient of eqn. (\ref{fit}) we find
$V_{r}$ = 1.5$\times 10^{4}$~cm~s$^{-1}$, implying that the average
aspect ratio and viscosity parameter for SXTs in outburst have orders
of magnitude such that
\begin{equation}
\alpha (H/R) \sim 0.01.
\end{equation}
A typical high--state value for $\alpha$ is of order 0.1, so this
result is consistent with the expected aspect ratio $H/R \sim 0.1$ of
the outbursting disc.

\subsection{Tra X--1 (=4U1524--62)}

The 1974 November X--ray outburst light curve of Tra X--1 is best fitted
with an exponential decay (see Fig. 3). There is a secondary maximum
present $\sim$ 52 days after the outburst peak; the peak observed flux is
$3.0\times 10^{-9}$~ergs~s$^{-1}$~cm$^{-2}$ and $\tau_{d}$=57.4 days.
Using equation (\ref{d}) below we estimate $D_{kpc}$=16 and 12 for the
neutron star and black hole cases respectively.

Using the lower limit to the distance we can estimate the absolute magnitude of
the secondary star. Given $V\sim 21$ and $A_{V}=2.4$ mags (CSL) we find
$M_{V}>$+3.2. The secondary star is either a main sequence star later
than F or a subgiant later than K5; the latter seems to be more
probable.

\subsection{SAX~J1808.4--3658}

The 1998 April X--ray outburst light curve of the X-ray pulsar/transient
SAX~J1808.4--3658 is best fitted with an exponential decay (see Fig. 6).
There is also a secondary maximum present $\sim$ 13 days after the
outburst peak. Given the observed peak flux $2.38\times
10^{-9}$~ergs~s$^{-1}$~cm$^{-2}$ (Gilfanov et al. 1998) and
$\tau_{d}$=8.4 days, using equation (\ref{d}) we estimate the distance to
be 5.5 kpc, assuming $f$=1.0. However, if $f$=0.5, then the we obtain 3.9
kpc which agrees very well with the distance estimate obtained from
assuming that the type $\sc i$ X-ray burst was Eddingtion limited (in't
Zand et al. 1998).

\section{Conclusions}

We have shown that the X--ray light curves of soft X--ray transients
are divided into exponential or linear according as their peak
luminosities lie above or below the critical value which can
maintain the outer disc in the hot state. This is as expected in the
model of King and Ritter (1998) for transient outbursts. We were also
able to estimate the radius and mass of the heated disc region. For
exponential declines these are consistent with the idea
that the disc is filled to a significant fraction of 
the maximum possible mass before
the outburst is triggered, presumably near its outer edge. This edge is 
found to be close to the circularization radius, as expected. 

For exponential decays the secondary maximum appears to occur one
irradiated--state viscous time after the start of the outburst, as
suggested by King and Ritter (1998). This result can be used to provide an
estimate of the distance to the system.

\section*{Acknowledgments}

ARK thanks PPARC for a Senior Fellowship.

\begin{figure*} 
\rotate[l]{\epsfxsize=500pt \epsfbox[00 00 700 750]{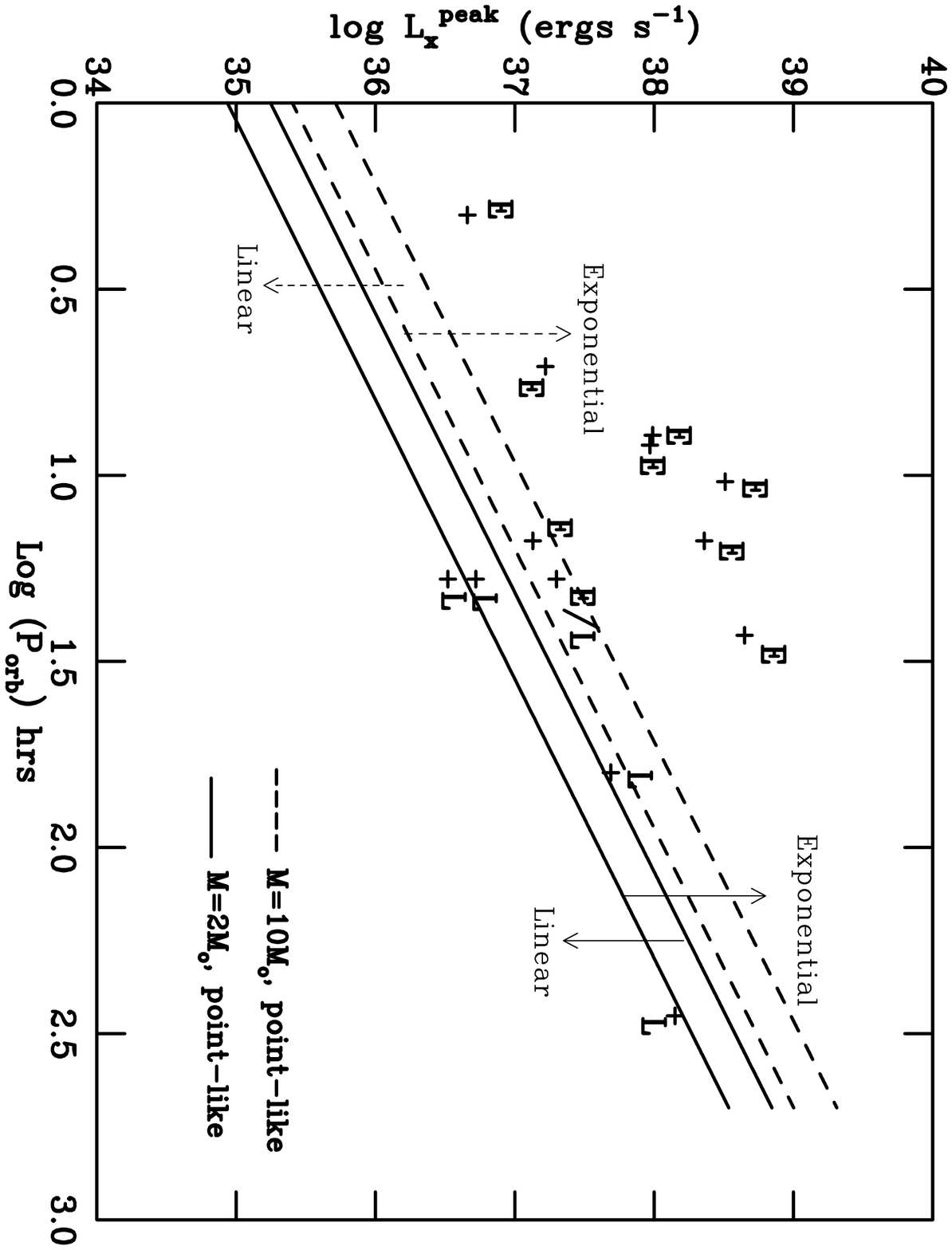}}
\caption{The critical luminosity needed to ionize the entire disc. The
dotted and solid lines show the critical luminosities for binaries with
total mass $M = 2M_{\odot}, 10M_{\odot}$, typical for neutron--star and
black--hole systems respectively. The critical luminosities are a factor
2 smaller for exponential decays than for linear ones because the disc
radii in these two cases are assumed to differ by a factor 0.7,
corresponding to the circularization radius and tidal radius
respectively. The systems given in Table 1 are plotted here, i.e. the
SXTs with known orbital periods and whose outburst decline shape can be
determined. In order of increasing orbital period, the SXTs shown are
SAX~J1808.4--3658, GRO~J0422+32, A0620--00, GS2000+25, GS1124--68, Cen
X--4, Aql X--1, 4U1543--47, GRO~J1655--40 and GRO~J1744--28. E and L
denote observed exponential and linear decays respectively. }
\end{figure*}

\begin{figure*} 
\epsfxsize=500pt \epsfbox[00 00 700 750]{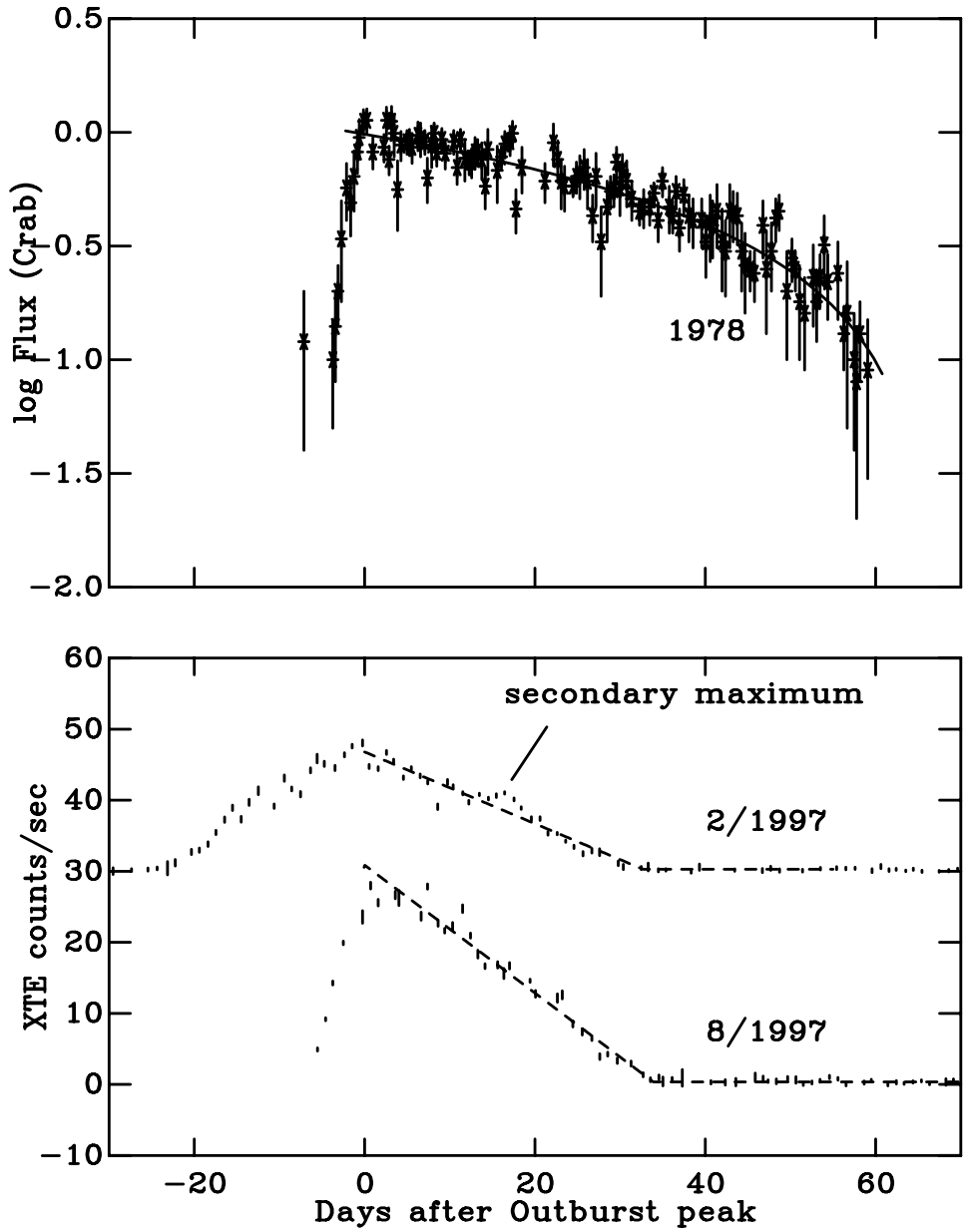}
\caption{The X--ray light curves of Aql X--1. The top panel shows the
1978 (Ariel 5/ASM) outburst. The bottom panel show the $RXTE$ February and
August 1997 outbursts. The February outburst has been shifted vertically by 10
counts/sec for clarity. We show on all the light curves linear fits to
the decay.}
\end{figure*}

\begin{figure*} 
\rotate[l]{\epsfxsize=500pt \epsfbox[00 00 700 750]{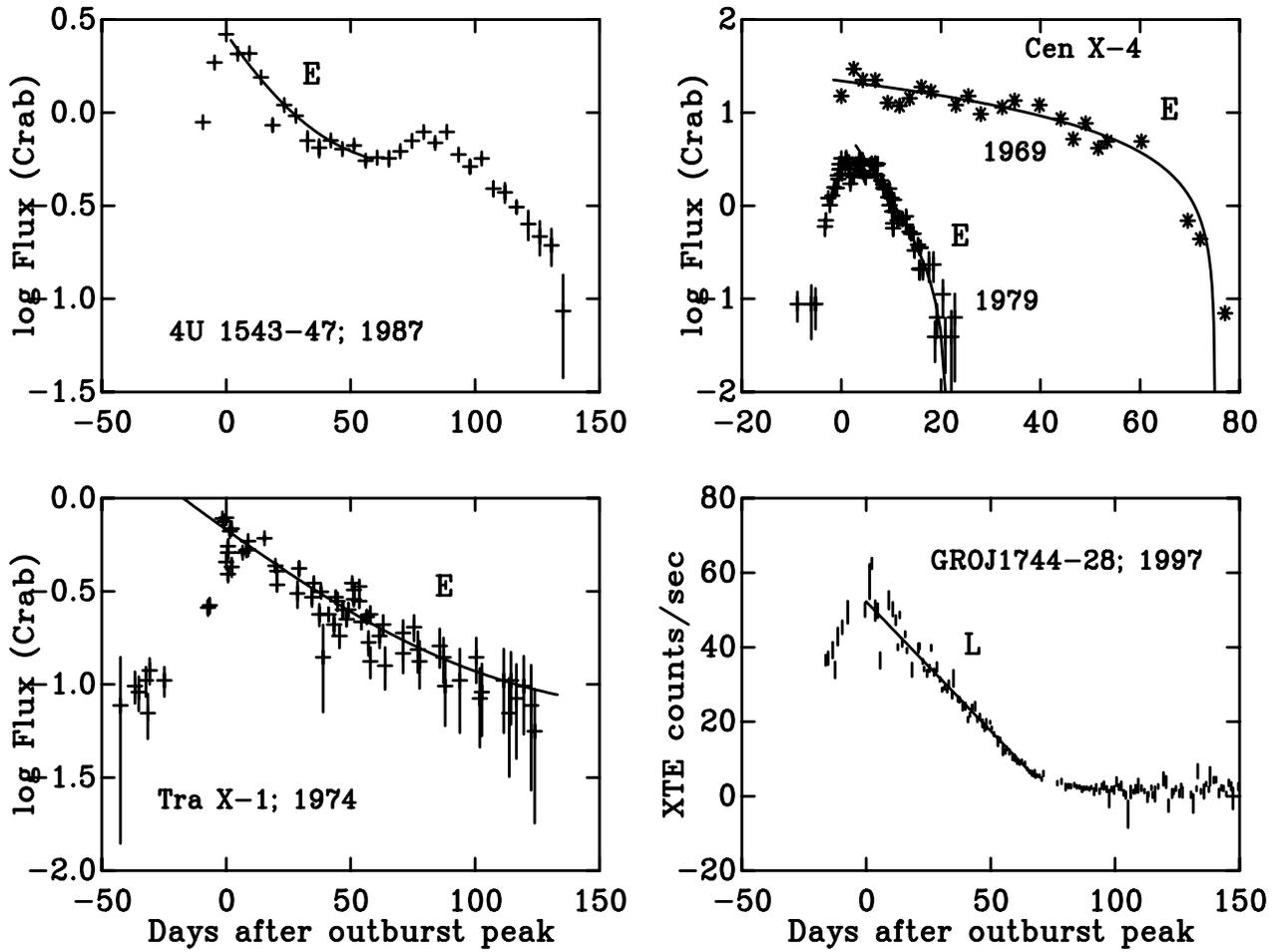}}
\caption{
X-ray light curves of SXTs. Top left: 1971 (Vela 5) X-ray outburst of
4U1543--47; top right: 1969 (Ariel 5/ASM) and 1979 (Vela 5) X-ray
outbursts of Cen X--4; bottom left: 1974 (Ariel 5/ASM) X-ray outburst
of 4U1524--62; bottom right: 1997 $RXTE$ X-ray outburst of GS1354--64. E
and L denote exponential and linear fits respectively. The data were
taken from CSL}
\end{figure*}

\begin{figure*} 
\rotate[l]{\epsfxsize=500pt \epsfbox[00 00 700 750]{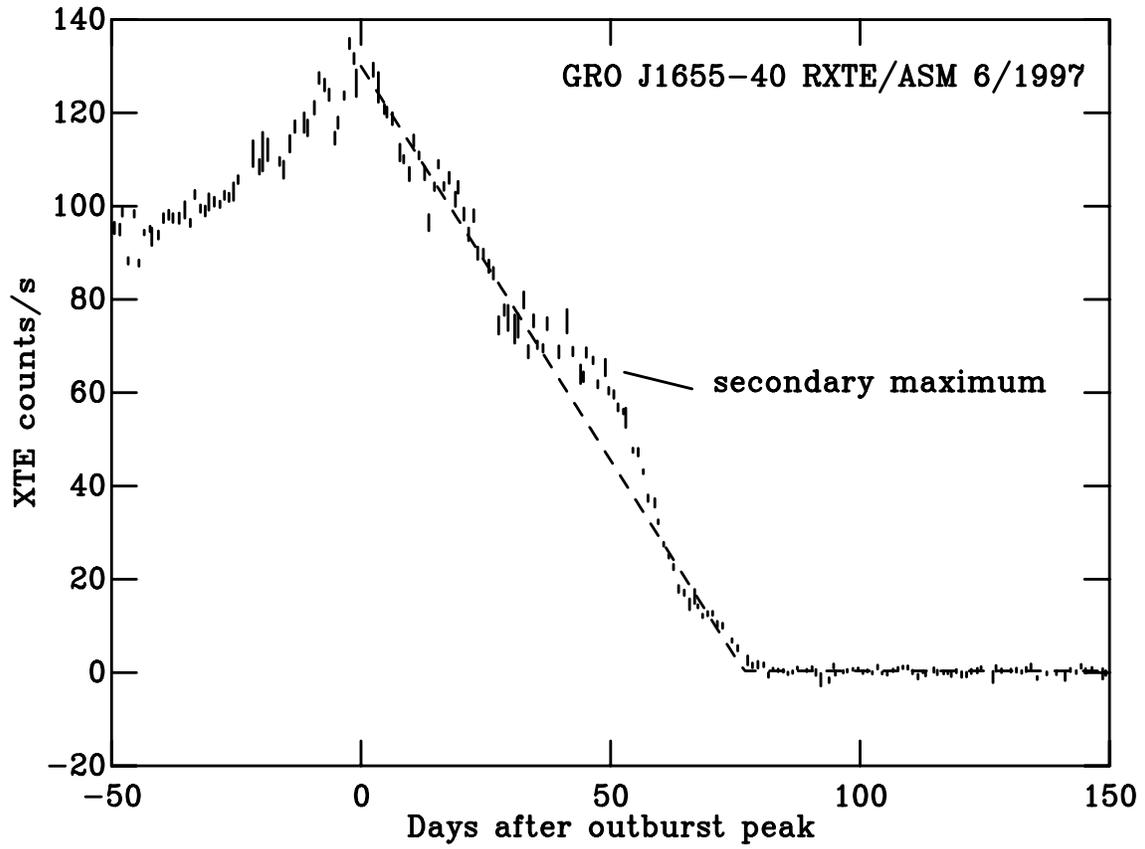}}
\caption{The $RXTE/ASM$ 1996 X-ray light curve GRO~J1655--40. The secondary
maximum can clearly be seen. A linear fit is also shown.}
\end{figure*}

\begin{figure*} 
\rotate[l]{\epsfxsize=500pt \epsfbox[00 00 700 750]{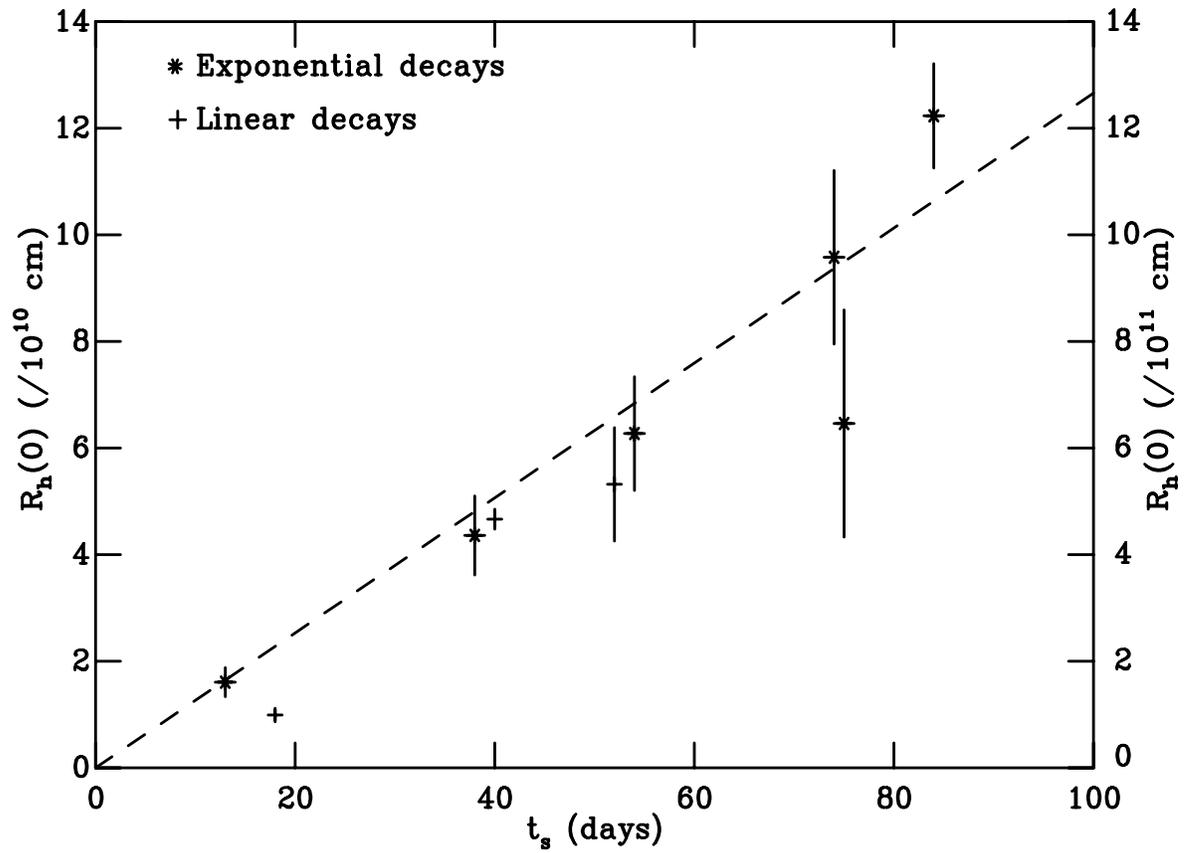}}
\caption{The radius of the heating front at the peak of the outburst
($R_{h}$) versus the time of the secondary maximum after the peak of the
outburst ($t_{s}$). The stars and crosses are data taken 
from exponential and linear decays respectively. 
The left and right ordinate refer to the
exponential and linear decay data points respectively. 
A linear fit to the exponential decays is also shown.}
\end{figure*}

\begin{figure*} 
\rotate[l]{\epsfxsize=500pt \epsfbox[00 00 700 750]{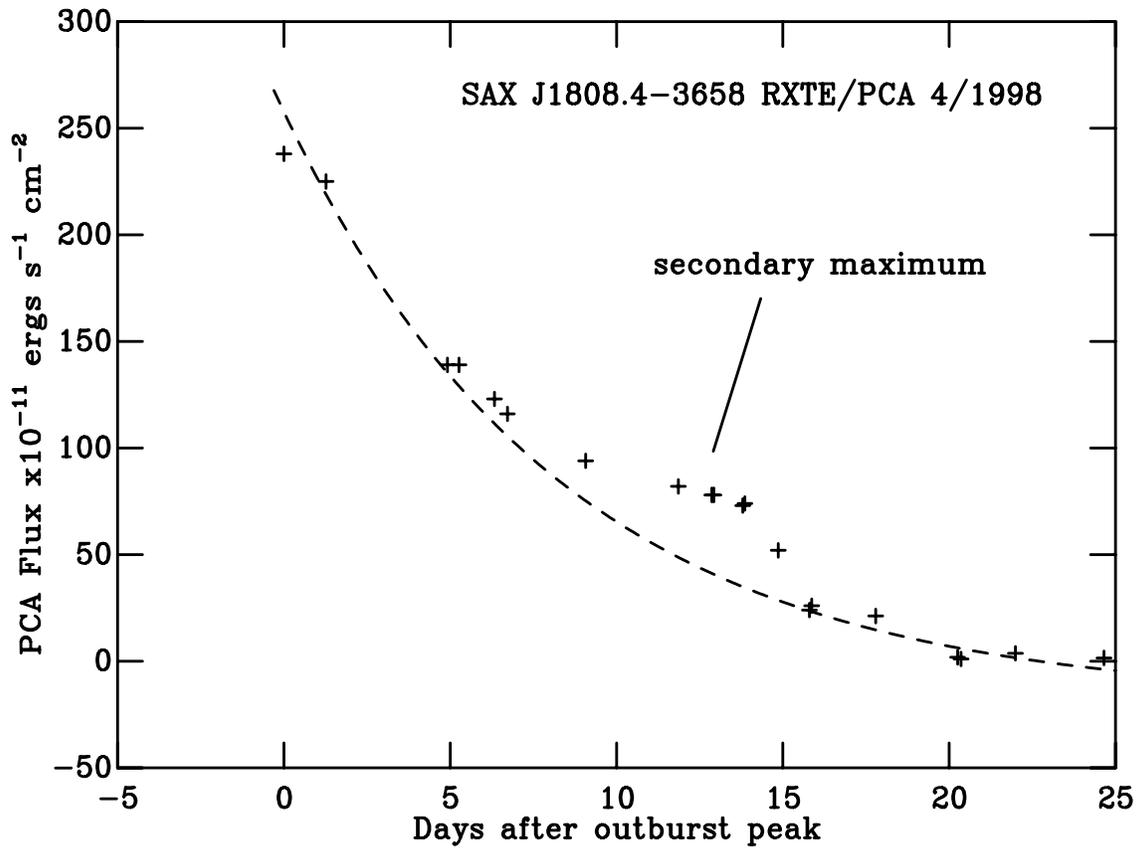}}
\caption{The $RXTE/PCA$ 1998 X-ray light curve SAX~J1808.4--3658
(Gilfanov et al. 1998). The secondary
maximum can clearly be seen. An exponential fit is also shown.}
\end{figure*}

\end{document}